\hsize=125mm
\vsize=185mm
\parindent=8mm
\frenchspacing
\overfullrule=0pt

\font \Large=cmbx10 scaled \magstep 2
\font \large=cmbx10 scaled \magstep 1


 


\def\Diff{\hbox{\rm Diff}}
\def\S1{\hbox{\rm S$^1$}}

\def\Vect{\hbox{\rm Vect}}

\def\ad{\hbox{\hbox{\rm ad}}}

\def\frac#1#2{{#1 \over #2}}

\def\pds#1,#2{\langle #1\mid #2\rangle} 
\def\f#1,#2,#3{#1\colon#2\to#3} 

\def\hfl#1{{\buildrel{#1}\over{\hbox to
12mm{\rightarrowfill}}}}

 
\null
\vskip1cm

\centerline{\Large  Coadjoint representation of Virasoro-type } 
\medskip 
\centerline{\Large  Lie algebras and differential operators}
\medskip
\centerline{\Large on  tensor-densities} 
\vskip1.8cm
\centerline{\large Valentin Yu. Ovsienko}
\bigskip
\centerline{Centre de Physique Th\'eorique}
\centerline{C.N.R.S.}
\centerline{Luminy -- Case 907}
\centerline{F-13288 Marseille Cedex 9}
\centerline{ France}
\centerline{email: Valentin.Ovsienko@cpt.univ-mrs.fr }

\vskip 0.8cm
\hfill{\it To my teacher Alexander Alexandrovich Kirillov}

\vskip 1cm

\centerline{\bf Abstract}
\bigskip
\noindent 
We discuss the geometrical nature of the coadjoint representation of the Virasoro algebra
and some of its generalizations.  The isomorphism of the coadjoint representation of the
Virasoro group to the $\Diff(S^1)$-action on the space of Sturm-Liouville operators was
discovered by A.A. Kirillov and G. Segal. This deep and fruitful result relates this topic
to the classical problems of projective differential geometry (linear differential
operators, projective structures on $S^1$ etc.) The purpose of this talk is to give a
detailed explanation of  the A.A. Kirillov method [14] for the geometric realization of
the coadjoint representation in terms of linear differential operators. Kirillov's method
is based on Lie superalgebras generalizing the Virasoro algebra. One obtains the
Sturm-Liouville operators directly from the coadjoint representation of these Lie
superalgebras. We will show that this method is universal. We will consider a few examples
of infinite-dimensional Lie algebras  and show that the Kirillov method can be applied to
them. This talk is purely expository: all the results are known.


\noindent Secondary: 17B65, 81R10. 

\vfill\eject
\noindent{\bf Table of Contents}
\bigskip
 
\halign {#&\quad #\quad &#\cr
\multispan 3 Introduction \hfill \cr
\noalign{\vskip2.5pt}
\hskip3.5pt 1 & \multispan 2   Coadjoint representation of Virasoro group 
and Sturm-Liouville operators; \hfill \cr 
\hskip3.5pt \phantom{1} & \multispan 2 Schwarzian derivative as a 1-cocycle\hfill \cr
\noalign{\vskip2.5pt}
\hskip3.5pt 2 & \multispan 2  Projectively invariant version of the Gelfand-Fuchs cocycle and 
of the \hfill\cr
\hskip3.5pt \phantom{2} & \multispan 2 Schwarzian derivative \hfill\cr 
\noalign{\vskip2.5pt}
\hskip3.5pt 3 & \multispan 2  Kirillov's method of Lie superalgebras \hfill\cr
\noalign{\vskip2.5pt} 
\hskip3.5pt 4 & \multispan 2  Invariants of coadjoint representation
of the Virasoro group \hfill\cr
\noalign{\vskip2.5pt}
\hskip3.5pt 5 & \multispan 2  Extension of the Lie algebra 
of first order linear differential operators on $S^1$  \hfill\cr
\hskip3.5pt \phantom{5} & \multispan 2 
and matrix analogue of the Sturm-Liouville operator \hfill\cr
\noalign{\vskip2.5pt}
\hskip3.5pt 6 & \multispan 2  Geometrical definition of the Gelfand-Dickey bracket 
and the relation to the  \hfill\cr
\hskip3.5pt \phantom{6} & \multispan 2 Moyal-Weil star-product \hfill\cr
\noalign{\vskip2.5pt}
\multispan 3 References \hfill\cr  }

\vskip0.7cm
{\Large Introduction}
\bigskip
\noindent
The coadjoint representation of infinite-dimensional
Lie groups and Lie algebras is one of the most interesting subjects
of Kirillov's orbit method.
Geometrical problems related to this subject link together
such fundamental domains as: symplectic and K\"ahler geometry,
harmonic analysis, integrable systems and many others.

The main purpose of this talk is to describe a ``geometrical picture''
due to Kirillov, for the coadjoint representation of the Virasoro
group and the  Virasoro algebra.
We will also consider
some of their generalizations.

\vskip 0,3cm

\noindent {\bf 1}. The Virasoro group is the unique (modulo equivalence)
nontrivial central extension of the group of diffeomorphisms of the circle.
The corresponding Lie algebra, called the Virasoro algebra,
 is defined as the unique (modulo equivalence)
nontrivial central extension of the Lie algebra of vector fields on $S^1$.
The coadjoint representation of the Virasoro group and the Virasoro
algebra was studied in pioneering works by A.A. Kirillov [13] 
and G. Segal [27]. Their result is as follows.

The dual space to the Virasoro algebra can be realized as the space
of Sturm-Liouville operators:

$$
L=c {{d^2} \over {dx^2}}+u(x)
\eqno{(1)}
$$

\noindent where $u(x+2\pi)=u(x)$ is a periodic function, 
$c\in{\bf R}$ (or ${\bf C}$) is a constant. The coadjoint 
representation of the Virasoro group coincides
with the natural action of
the group of diffeomorphisms of $S^1$ on the space of operators 
(1).

This realization gives a geometric interpretation of the
coadjoint representation of the Virasoro group (and the Virasoro algebra).
It relates the coadjoint representation of the Virasoro group
to classical works on
differential operators and projective differential geometry
[29],[3]. 
The main object in this theory which
links together its different parts,
is the classical Schwarzian derivative.
It appears (in the Virasoro context) 
as a 1-cocycle on the group of diffeomorphisms of $S^1$
with values in the coadjoint representation.

\vskip 0,3cm

\noindent {\bf 2}. The realization of the coadjoint representation of the Virasoro
group was first discovered as a simple coincidence.
Soon after, A.A. Kirillov has suggested a systematic method
using Lie superalgebras (see [14]).
He considered two Lie superalgebras (called now Ramond and Neveu-Schwarz superalgebras)
containing the Virasoro algebra as the even part.
Sturm-Liouville operator appears in the coadjoint action of
Ramond and Neveu-Schwarz superalgebras.

To my knowledge, Kirillov's method is the only known way to
obtain the Sturm-Liouville operators (in an automatic way) 
directly from the coadjoint representation.
This makes this method particularly useful for generalizations.
However, for a long time, Kirillov's method has not been tested
in any other case then for the Virasoro algebra.

\vskip 0,3cm

\noindent {\bf 3}. We consider two different generalizations of the above geometrical picture.

{\bf A}. There exist series of infinite-dimensional
groups, Lie algebras and Lie superalgebras generalizing the Virasoro
group and the Virasoro algebra (see [21], [25]).
Geometrical realization of the coadjoint representation leads to interesting
generalization of the Sturm-Liouville operator and projective structures.

\vskip 0,3cm

{\bf B}. The space of higher order linear differential operators has an
interesting structure of infinite-dimensional Poisson manifold
with respect to the Adler-Gelfand-Dickey Poisson bracket.
This Poisson structure is related to so-called $W$-algebras
and is very popular in Mathematical Physics (see e.g. 
[28]).
We will discuss the relations of the Adler-Gelfand-Dickey bracket to the 
$\Diff(S^1)$-module structure on the space of 
linear differential operators on $S^1$ (studied by classics,
see [29], [3]).
Following [24], we show that the Adler-Gelfand-Dickey
Poisson structure can be defined in terms of the Moyal-Weyl
star-product.

\vskip 0,3cm

\noindent {\bf 4}. An important point, common for all known examples
is the following ``tensor sense''.
Arguments of differential operators are considered as 
tensor-densities on $S^1$. The well-known classical
example is the Sturm-Liouville operator, (1) acting from the space
of $-1/2$-densities to the space of $3/2$-densities.
This defines a natural $\Diff(S^1)$-action on the space of
differential operators and intrinsically contains all the information
about the related algebraic structures.

Remark here that the structure of the module over the group of diffeomorphisms
on the space of linear differential operators on a manifold
was studied in a series of recent papers [5], [20], [6].

\vskip 0,5cm

\noindent {\bf Acknowledgments}. I am grateful to 
A.A. Kirillov for his constant
help and to Ch. Duval,
L. Guieu, P. Marcel and C. Roger for collaboration
and
numerous stimulating discussions.

\vfill\eject

\item{\Large 1}{\Large Coadjoint representation of Virasoro group 
and 

Sturm-Liouville operators;
Schwarzian derivative as a 1-cocycle}
\bigskip
\noindent
This introductory section is based on the articles of A.A. Kirillov
[13], [14] and G. Segal [27]. 
We will give the definition of the Virasoro group
and the Virasoro algebra and prove the following result.
\medskip
\noindent
{\bf Theorem 1.}{\sl The coadjoint representation
of the Virasoro group is naturally isomorphic, i.e., isomorphic as a 
module over the group of diffeomorphisms, to the space of Sturm-Liouville
operators.}
\medskip
\noindent The classical Schwarzian derivative appears as a 1-cocycle on
the group of diffeomorphisms with values in the coadjoint representation.
\bigskip
\noindent

{\large 1.1 Virasoro group and Virasoro algebra}

\medskip
\noindent
Consider the Lie algebra $\Vect (S^1)$ of smooth vector fields on the circle:
$$
X=X(x)\frac{d}{dx},
$$
where $X(x+2\pi)=X(x)$.
The commutator in $\Vect (S^1)$ is given by the formula:
$$
[X(x)\frac{d}{dx},Y(x)\frac{d}{dx}]=
(X(x)Y^{\prime}(x)-X^{\prime}(x)Y(x))\frac{d}{dx},
$$
where $X'=dX/dx$.

\vskip 0,3cm

\noindent {\bf Definition 1.1}. The {\it Virasoro algebra} is 
the unique (up to isomorphism) non-trivial central
extension of $\Vect (S^1)$. It is given by the {\it Gelfand-Fuchs cocycle}:
$$
 \omega(X(x)\frac{d}{dx},Y(x)\frac{d}{dx})=\frac{1}{2}\int_{0}^{2\pi}
\left| 
\matrix{
X^{\prime}(x)&Y^{\prime}(x) \cr
X^{\prime\prime}(x)&Y^{\prime\prime}(x)
}
\right|dx .
$$

\vskip 0,3cm

\noindent The Virasoro algebra is, therefore a Lie algebra on the space
$\Vect (S^1)\oplus\bf R$ defined by the commutator
$$
[(X,\alpha),(Y,\beta)]=([X,Y]_{\rm{Vect}(S^1)},\omega(X,Y)),
$$
where $\alpha$ and $\beta\in {\bf R}$ are elements of the center.

\vskip 0,3cm

\noindent One easily checks (using integration by part) the 
2-cocycle condition:
$$
\omega(X,[Y,Z])+\omega(Y,[Z,X])+\omega(Z,[X,Y])=0,
$$
equivalent to the Jacobi identity
for the Virasoro algebra.

\vskip 0,3cm

\noindent {\bf Remark}. The defined Lie algebra has been discovered 
by I.M. Gelfand and D.B. Fuchs [8]
and was later rediscovered in the physics literature.

\vskip 0,3cm

\noindent Consider the group $\Diff^+(S^1)$
of diffeomorphisms of the circle preserving its orientation:
$
x\mapsto f(x),
$
where $x(\hbox{mod}\;2\pi)$ is a parameter on $S^1$, $f(x+2\pi)=f(x)+2\pi$.

\vskip 0,3cm

\noindent {\bf Definition 1.2}. The {\it Virasoro group} is 
the unique (up to isomorphism) non-trivial central
extension of $\Diff^+(S^1)$. It is given by so-called 
{\it Thurston-Bott cocycle}
(see [2]):
$$
 B(f,g)=\int_{0}^{2\pi}\log((f \circ g)')d\log(g ')
$$

\vskip 0,3cm

\noindent By definition, the Virasoro group is given 
by the following product on $\Diff^+(S^1)\times{\bf R}$:
$$
(f,\alpha)(g,\beta)=(f\circ g,\beta+B(f,g)).
$$
Associativity of this product is equivalent to 
the condition: $B(f,g\circ h)+B(g,h)=B(f\circ g,h)$
which means that $B$ is a 2-cocycle.

\vskip 0,3cm

\noindent {\bf Notation}. Let us denote the Virasoro algebra by
$vir$.

\bigskip
\noindent
{\large 1.2 Regularized dual space}
\medskip
\noindent The dual space to the Virasoro algebra:
$$
vir^*\cong\Vect(S^1)^*\oplus{\bf R}
$$ 
consists of pairs:
$(u,c)$ where $u$ is a distribution on $S^1$ and $c\in{\bf R}$.
Following A.A. Kirillov (cf. [13]),
we will consider only the {\it regular part}
of the dual space, $vir^*_{\rm{reg}}$
corresponding to distributions given by smooth functions.
In other words, 
$
vir^*_{\rm{reg}}\cong C^{\infty}(S^1)\oplus\bf R.
$

Geometrically, objects dual to vector fields on $S^1$ have the sense of
quadratic differentials:
$u=u(x)(dx)^2$ (cf. [13], formulae (2) and (4) below).
Note, that as a vector space the space of quadratic differentials
${\cal F}_2\cong C^{\infty}(S^1)$.
One obtains the following realization:
$$
vir^*_{\rm{reg}}={\cal F}_2(S^1)\oplus\bf R
$$
with the pairing:
$$
\langle (u(x)(dx)^2,c),(X(x)\frac{d}{dx},\alpha)\rangle=
\int_{0}^{2\pi}u(x)X(x)dx+c\alpha.
$$
Let us calculate the coadjoint action of the Virasoro algebra and of the
Virasoro group on space $vir^*_{\rm{reg}}$.

\bigskip
\noindent
{\large 1.3 Coadjoint representation of the Virasoro algebra}
\medskip
\noindent
Let us recall the definition.
The coadjoint representation of a Lie algebra $g$
is the action on its dual space $g^*$, defined by:
$$
\langle ad^*_X(\mu),Y\rangle=
-\langle \mu,[X,Y]\rangle,
$$
for every $X\in g$ and $\mu\in g^*$.

\vskip 0,3cm

\noindent 
The coadjoint action of the Virasoro algebra preserves the regular part
of the dual space. 
\medskip
\noindent
{\bf Lemma 1.3.} {\sl The coadjoint action of the Virasoro algebra
on the regular part of its dual space is given by the formula
$$
ad^*_{(X(x)\frac{d}{dx},\alpha)}(u(x)(dx)^2,c)=
(L_X(u)-c\cdot X'''(x)(dx)^2,\;0)
\eqno{(2)}
$$
where $L_X(u)$ is the Lie derivative of a quadratic differential $u$:
$$
L_X(u)=
(X(x)u'(x)+2X'(x)u(x))(dx)^2.
$$  }

\noindent 
{\bf Proof}. By definition 
$$
\langle  ad^*_{(X(x)\frac{d}{dx},\alpha)}(u(x),c),
\;(Y(x)\frac{d}{dx},\beta)\rangle=
-\langle (u(x),c),\;\bigg[(X(x)\frac{d}{dx},\alpha),
(Y(x)\frac{d}{dx},\beta)\bigg]\rangle $$
$$=-\int_0^{2\pi}u(XY'-X'Y)dx-\frac{c}{2}\int_0^{2\pi}(X'Y''-X''Y')dx. 
$$
Integrating by parts, one obtains the expression:
$$
\int_0^{2\pi}(Xu'+2X'u-cX''')Ydx.
$$
The lemma follows.

\vskip 0,3cm
\noindent 
Note, that the coadjoint action of the Virasoro algebra is in fact,
a $\Vect (S^1)$-action (the center acts trivially).

\vskip 0,3cm

\noindent 
{\bf Remarks}. (a) The case $c=0$ corresponds to the coadjoint action of
$\Vect(S^1)$ (without central extension). This is just the natural
$\Vect(S^1)$-action by the Lie derivative
on the space ${\cal F}_2$ of quadratic differentials.

\vskip 0,3cm

(b) The linear map:
$$
s:X(x)\frac{d}{dx}\;\mapsto \;X'''(x)(dx)^2
\eqno{(3)}
$$
is a {\it 1-cocycle} on $\Vect(S^1)$ with values in ${\cal F}_2$.
It satisfies the relation: 
$L_X(s(Y))-L_Y(s(X))=s([X,Y])$.

\bigskip
\noindent
\item{\large 1.4}{\large The coadjoint action of Virasoro group 
and Schwarzian derivative}
\medskip
\noindent
The coadjoint action of the  Virasoro group
on the regular part of $vir^*$ is the ``group version'' of the
$\Vect(S^1)$-action (2). As in the case of the Virasoro
algebra, the center acts trivially and therefore, the coadjoint representation of
the  Virasoro group is just a $\Diff^+(S^1)$-representation.

It is clear that this action is
of the form:
$$
Ad^*_{f^{-1}}(u,c)=
(u\circ f-c\cdot S(f),\;c)
\eqno{(4)}
$$
where 
$$
u\circ f=u(f(x))(df)^2
$$
is the natural $\Diff^+(S^1)$-action on ${\cal F}_2$
and $S$ is some 1-cocycle on $\Diff^+(S^1)$ with values in ${\cal F}_2$.
Indeed, this action corresponds to the coadjoint action of the Virasoro
algebra (2).

\vskip 0,3cm

\noindent 
The explicit formula for the 1-cocycle $S$ was calculated in [13]
and [27]. 
\medskip
\noindent
{\bf Proposition 1.4 [13].} {\sl The coadjoint action 
of the Virasoro group on the regular dual space $vir^*_{\rm{reg}}$
is defined by the 1-cocycle
 $$
S(f)=\big(f'''/f'-\frac{3}{2}(f''/f')^2\big)(dx)^2
\eqno{(5)}
$$   }

\vskip 0,3cm

\noindent {\bf Notation}. The cocycle (5) is called the
{\it Schwarzian derivative}.

\vskip 0,3cm

\noindent An elegant proof of the formulae (4), (5) directly from
the definition of the coadjoint representation can be found in [13]. 

\vskip 0,3cm

\noindent One can also deduce these formulae from (2).
To do this, it is sufficient to check the following two properties:

\vskip 0,3cm

\noindent 
(a) The formula (4) indeed defines an action of $\Diff^+(S^1)$:
$$
Ad^*_f\circ Ad^*_g=Ad^*_{f\circ g}.
$$
This follows from the well-known property of the Schwarzian derivative:
$$
S(f\circ g)=S(f)\circ g+S(g).
$$
Which means that the mapping $f\mapsto S(f)$
is a {\it 1-cocycle} on $\Diff^+(S^1)$ with values in $vir^*$.

\vskip 0,3cm

\noindent 
(b) The action (2) is the infinitesimal version of (4).

\bigskip
\noindent
{\large 1.5 Space of Sturm-Liouville equations
as a $\Diff^+(S^1)$-module}
\medskip
\noindent
It turns out that 
the formulae (2) and (4) 
has already been known to
classics for a
long time before the discovering of the Virasoro algebra.

Consider the (affine) space of Sturm-Liouville operators (1).
There exists a natural $\Diff^+(S^1)$-action
on this space  (cf. [3],[29]).
It turns out that this action coincides with the coadjoint action (4).

\vskip 0,3cm

\noindent 
{\bf Definition 1.5}. 
Consider a one-parameter family of actions of $\Diff^+(S^1)$ on
the space of functions on $S^1$:
$$
g^*_{\lambda}a=a\circ g^{-1}\bigg((g^{-1})'\bigg)^{\lambda}
\eqno{(6)}
$$

\vskip 0,3cm

\noindent 
{\bf Notation}. Denote ${\cal F}_{\lambda}$ the $\Diff^+(S^1)$-module 
structure 
(6) on space $C^{\infty}(S^1)$.

\vskip 0,3cm

\noindent {\bf Remark}. 
Geometrically speaking, $a$ has the sense of tensor-density 
of degree $\lambda$ on $S^1$:
$$
a=a(x)(dx)^{\lambda}
$$
and the action (6) becomes simply
$g^*a=a\circ g^{-1}$.

\vskip 0,3cm

\noindent 
Let us look for a  $\Diff^+(S^1)$-action
on the space of Sturm-Liouville operators
in the form:
$g^*_{\lambda\mu}(L)=g^*_{\lambda}\circ L\circ(g^*_{\mu})^{-1}$
for some $\lambda,\mu$.
It is easy to check that this formula preserves the space of Sturm-Liouville operators
(this means, the differential operator $g^*_{\lambda\mu}(L)$ is 
again an operator of the form (1))
if and only if $\lambda=3/2, \mu=-1/2$.

\vskip 0,3cm

\noindent 
{\bf Definition 1.6}. The action of group $\Diff^+(S^1)$ 
on the space of differential operators (1)
is defined by:
$$
g^*(L):=g^*_{\frac{3}{2}}\circ L\circ(g^*_{-\frac{1}{2}})^{-1}
\eqno{(7)}
$$
(see [29],[3]).

\vskip 0,3cm

\noindent 
In other words, Sturm-Liouville operators are considered as acting
on tensor-densities:
$$
L:{\cal F}_{-1/2}\to{\cal F}_{3/2}.
$$

\noindent 
The following statement has already been known to classics.
\medskip
\noindent
{\bf Proposition 1.7.} {\sl The result of the action (7)
is again a Sturm-Liouville operator:
$g^*(L)=c\cdot d^2/dx^2+u^g$ with the potential
$$
u^g=u\circ g^{-1}\bigg((g^{-1})'\bigg)^2+\frac{c}{2}\cdot S(g^{-1}).
$$  }

\noindent 
{\bf Proof.} Straightforward.
\bigskip
\noindent
{\large 1.6 The isomorphism}
\medskip
\noindent
The last formula coincides with the coadjoint action of the 
Virasoro group (4)
(up to the multiple $-1/2$ in the last term).
This remarkable coincidence shows that the space of Sturm-Liouville operators
is isomorphic as a $\Diff^+(S^1)$-module to the coadjoint representation.

The isomorphism is given by the formula:
$$
(u,c)\;\;\longmapsto \;\;-2c\cdot d^2/dx^2+u(x)
\eqno{(8)}
$$

\vskip 0,3cm

\noindent Theorem 1 is proven.

\medskip
\noindent
{\large 1.7 $\Vect(S^1)$-action on the space of Sturm-Liouville operators}
\medskip
\noindent
The infinitesimal version of the $\Diff^+(S^1)$-action 
on the space of Sturm-Liouville operators
is given by the commutator with the Lie derivative:
$$
\ad L_X(L):= L_X^{3/2}\circ L-L\circ L_X^{-1/2}
\eqno{(9)}
$$
where $L_X^{\lambda}$ is the operator of $\Vect(S^1)$-action on 
${\cal F}_{\lambda}$.
In other words, $L_X^{\lambda}$ is the operator
of Lie derivative on the space of tensor-densities of degree 
$\lambda$:
$$
L_X^{\lambda} = X {d \over {dx}} + \lambda \cdot X'
\eqno{(10)}
$$
\medskip
\noindent
{\bf Proposition 1.8.} {\sl The result of the action (9)
is a scalar operator of multiplication by:
$$
\ad L_X(L)=Xu'+2X'u-c\cdot X'''
$$  }
\medskip
\noindent
{\bf Proof}. This formula can be proven by simple direct calculations.

\vskip 0,3cm

\noindent 
The proposition follows also from the isomorphism (8).
Indeed, the coadjoint action of $X$ associates to the pair
$(u(x)(dx)^2,c)$ the expression $((Xu'+2X'u-c\cdot X''')(dx)^2,0)$
corresponding to the scalar operator.

\vskip 0,3cm

\noindent {\bf Remarks}. 
(a) The operator $L$ (and therefore $\ad L_X(L)$)
maps from ${\cal F}_{-1/2}$ to ${\cal F}_{3/2}$.
This means that the scalar operator  $\ad L_X(L)$
is rather an operator of multiplication by a tensor-density
of degree 2 (a quadratic differential) then by a function:
$\ad L_X(L)\in{\cal F}_2$.

\vskip 0.1cm

(b) The formula (9) corresponds to the $\Diff^+(S^1)$-action
(7). 
However, the ``point of view'' of Lie algebras is much more universal:
it works in the case (of Lie algebras more general then the Virasoro algebra)
when there is no corresponding Lie group and there is no analogue of
the formula (7).

\bigskip
\noindent
\item{\Large 2} {\Large Projectively invariant version of the Gelfand-Fuchs cocycle and 
of the Schwarzian derivative}
\bigskip
\noindent
In this section we follow G. Segal (see [27]).

As it was mentioned, the Virasoro algebra is the unique nontrivial central extension
of $\Vect(S^1)$. However, 
the 2-cocycle on $\Vect(S^1)$ defined this extension can be chosen in different ways
(up to a coboundary). 
There exists a unique way to define the Gelfand-Fuchs cocycle and the cocycles
(3),(5) such that they are projectively invariant.

\vskip 0,3cm

\noindent 
Consider the subalgebra of $\Vect(S^1)$ generated by the vector fields:
$$
\frac{d}{dx},
\;\;\;\sin x\frac{d}{dx},
\;\;\;\cos x\frac{d}{dx}.
\eqno (11)
$$
It is isomorphic to $sl_2({\bf R})$ and the corresponding Lie group is $PSL(2,{\bf R})$
acting on $S^1\;(\cong{\bf RP}^1)$ by {\it projective transformation}.

\bigskip
\noindent
{\large 2.1 Modified Gelfand-Fuchs cocycle}
\medskip
\noindent
Consider the following ``modified'' Gelfand-Fuchs cocycle
on $\Vect(S^1)$:
$$
 \bar\omega(X(x)\frac{d}{dx},Y(x)\frac{d}{dx})=\int_{0}^{2\pi}
(X^{\prime\prime\prime}+X^{\prime})Ydx
\eqno (12)
$$
\vskip 0,3cm

\noindent 
It is clear that this cocycle is cohomologous to the Gelfand-Fuchs cocycle
and therefore, the corresponding central extension is isomorphic to the Virasoro algebra.
Indeed, the additional term in (12) is a coboundary: the functional
$$
\int_{0}^{2\pi}X^{\prime}Ydx={1\over2}\int_{0}^{2\pi}(X^{\prime}Y-XY^{\prime})dx
$$
depends only on the commutator of $X$ and $Y$.

The cocycle (12) is
{\it $sl_2$-equivariant}. This means, 
$$
 \bar\omega([Z,X],Y)+ \bar\omega(X,[Z,Y])=0
$$
for every $X,Y\in \Vect(S^1)$ and $Z\in sl_2$.

\medskip
\noindent
{\bf Proposition 2.1.} {\sl The cocycle (12) is the unique (up to a constant)
$sl_2$-equivariant 2-cocycle on $\Vect(S^1)$. }

\noindent {\bf Proof}. Let $\widetilde\omega$ be a
$sl_2$-equivariant 2-cocycle on $\Vect(S^1)$.
The equivariance condition is equivalent to:
$$
\widetilde\omega(X,Z)\equiv0,\;\;\;Z\in sl_2
$$
Indeed, since $\widetilde\omega$ is a cocycle, one has:
$
\widetilde\omega([Z,X],Y)+
\widetilde\omega([X,Y],Z)+
\widetilde\omega([Y,Z],X)=0
$
The $sl_2$-equivariance condition gives now: $\widetilde\omega([X,Y],Z)=0$
for every $X,Y\in \Vect(S^1)$ and $Z\in sl_2$.
But the commutant in $\Vect(S^1)$ coincides with $\Vect(S^1)$.

The Gelfand-Fuchs theorem (see [8]) states that 
$H^2(\Vect(S^1))={\bf R}$, and therefore,
every nontrivial cocycle is proportional to
the Gelfand-Fuchs cocycle up to a coboundary.
One has:
$$
\widetilde\omega=\kappa\omega+b,
$$
where $b$ is a coboundary: $b(X,Y)=\langle u,[X,Y]\rangle$
for some $u\in \Vect(S^1)^*$.

The $sl_2$-equivariance condition means that $b(X,Z)=0$
for $Z\in sl_2$ and an arbitrary $X\in\Vect(S^1)$.
This implies $u=0$.

\bigskip
\noindent
{\large 2.2 Modified Schwarzian derivative}
\medskip
\noindent
It is easy to check that the modified action of $\Vect(S^1)$ on $vir_{\rm{reg}}$
is as follows:
$$
\overline{ad}^*_{X\frac{d}{dx}}(u(dx)^2,c)=
(L_X(u)-c\cdot(X'''+X')(dx)^2,\;0)
\eqno (13)
$$

\noindent 
The modified $\Diff(S^1)$-action on $vir_{\rm{reg}}$ is:
$$
\overline{Ad}^*_{f^{-1}}(u,c)=
(u\circ f-c\cdot \bar S(f),\;c)
\eqno (14)
$$
where $\bar S(f)$ is the following (modified) Schwarzian derivative:
$$
\bar S(f)=\bigg(\frac{f'''}{f'}-
\frac{3}{2}\left(\frac{f''}{f'}\right)^2
+\frac{1}{2}(f'^2-1)\bigg)(dx)^2
\eqno (15)
$$
The 1-cocycles $S$ and $\bar S$ on $\Diff^+(S^1)$ are cohomologous.

\vskip 0,3cm

\noindent {\bf Remarks}. 
(a) The following amazing fact often leads to confusion.
Take the affine parameter $t=\hbox{tg}(x/2)$.
Then, the modified Schwarzian derivative $\bar S$ is
given by the expression:
$\bar S(f(t))=\buildrel{\dots}\over{f}/\dot f-(3/2)(\ddot f/\dot f)^2$,
where $\dot f=df/dt$. 
This expression coincides with the formula (5) for $S$
(but $\bar S\not=S$).

\vskip 0,3cm

(b) The $PSL_2$-equivariant Schwarzian (15) has been
considered in [27] (see also [17]).

\bigskip
\goodbreak
\noindent
{\large 2.3 Energy shift}
\medskip
\noindent
The projectively invariant coadjoint action 
corresponds to another realization of the dual space to the Virasoro algebra 
as the space of
Sturm-Liouville operators.
 The map
$$
(u,c)\;\longmapsto \;-2c\frac{d^2}{dx^2}+u(x)+\frac{c}{2}
\eqno (16)
$$
is an isomorphism of $\Diff^+(S^1)$-module (16)
and the module of Sturm-Liouville operators.

\vskip 0,3cm

\noindent 
Indeed, the fact that the quantity $U=(u(x)+c/2)(dx)^2$ 
transforms according to the formulae
(13) and (14), means that the quadratic differential 
$u(x)(dx)^2$ transforms under the $\Diff^+(S^1)$-action via the formulae
(2) and (4).

\bigskip
\noindent
{\large 2.4 Projective structures}
\medskip
\noindent
Let us recall well-known definitions.

\vskip 0,3cm

\noindent 
An atlas $(U_i,t_i)$ on $S^1$ is called a {\it projective atlas}
if the coordinate transformations $t_j\circ t_i^{-1}$ are linear-fractional
functions. Two atlas' are called equivalent if their union is again a projective atlas.
A class of equivalent projective atlas' is called a {\it projective structure} on $S^1$.

Every projective structure on $S^1$ defines a (local) action of the Lie
algebra $sl_2({\bf R})$ generated by the vector fields
$$
\frac{d}{dt},\;\;\;\;t\frac{d}{dt},\;\;\;\;(t)^2\frac{d}{dt},
$$
where $t=t_i$ is a local coordinate of the projective structure.
This action is invariant under the linear-fractional transformations of $t_i$.

\vskip 0,3cm

\noindent {\bf Remark}. This $sl_2$-action coincides with the action (11)
for the angular parameter $x=\hbox{arctg}(t)$.

\vskip 0,3cm

\noindent There exists a natural isomorphism between the space of 
Sturm-Liouville operators
and the space of projective structures on $S^1$.
Given a Sturm-Liouville operator (1),
consider the corresponding differential equation:
$c\cdot\phi''+u(x)\phi=0$. Local coordinates of projective structure
associated to this operator are defined as functions of two independent
solutions:
$$
t=\frac{\phi_1}{\phi_2},
$$
on an interval with $\phi_2\not=0$.
An important remark is that for the local coordinate $t$, the potential of the
Sturm-Liouville operators is identically zero:
$$
c\frac{d^2}{dx^2}+u(x)\;=\;c\frac{d^2}{dt^2}.
$$

\vskip 0,3cm

\noindent 
A beautiful definition of the corresponding $sl_2$-action was proposed by
A.A. Kirillov (see [14]). It is given by products of solutions:
the generators are as follows:
$$
\phi_1^1,\;\;\;\;\phi_1\phi_2,\;\;\;\;\phi_2^1
$$
Note, that the solutions are $-1/2$-tensor-densities, therefore
their product is a vector field.
Use $W(\phi_1,\phi_2)=\phi_1\phi_2'-\phi_1'\phi_2=\hbox{const}$ to
verify that the these vector fields indeed generate a $sl_2$-subalgebra.

\bigskip
\noindent
\item{\Large 3} {\Large Kirillov's method of Lie superalgebras}
\bigskip
\noindent
The ``mysterious'' coincidence between the coadjoint representation
of the Virasoro group and the natural $\Diff^+(S^1)$-action on the 
space of Sturm-Liouville operators 
(Theorem 1) was explained by A.A. Kirillov in [14], where furthermore
an algebraic explanation of the fact that Sturm-Liouville operators 
act on tensor-densities is given.
It is an amazing fact, that the natural interpretation of these
geometrical results uses Lie superalgebras.

\bigskip
\noindent
{\large 3.1 Lie superalgebras}
\medskip
\noindent
A Lie superalgebra is a ${\bf Z}_2$-graded algebra
$$
g=g_0\oplus g_1
$$
with the multiplication called the commutator
satisfying two conditions:
\hfill\break
(1) superized skew symmetry: $[X,Y]+(-1)^{\widetilde X\widetilde Y}[Y,X]=0$,
\hfill\break
(2) superized Jacobi identity:
$$
(-1)^{\widetilde X\widetilde Z}[X,[Y,Z]]+
(-1)^{\widetilde X\widetilde Y}[Y,[Z,X]]+
(-1)^{\widetilde Y\widetilde Z}[Z,[X,Y]]=0,
$$
where $\widetilde X$ is the degree
($\widetilde X=0$ for $X\in g_0$ and $\widetilde X=1$ for $X\in g_1$).

\vskip 0,3cm

\noindent 
The simplest properties of Lie superalgebras are as follows:

\noindent (a) $g_0\subset g$ is a Lie subalgebra,

\noindent (b) $g_0$ acts on $g_1$ via
the commutator: if
$X\in g_0$ and $\xi\in g_1$ is $[X,\xi]\in g_1$

\noindent (c)
the product of $\xi,\eta\in g_1$ called an {\it anticommutator}
is defined by a symmetric bilinear map
$$
[\;,\;]_+:g_1\otimes g_1\to g_0.
$$

\bigskip
\noindent
{\large 3.2 Ramond and Neveu-Schwarz superalgebras}
\medskip
\noindent
Consider the space ${\cal F}_{-1/2}$ of $-1/2$-tensor densities on $S^1$:
$
\phi=\phi(x)(dx)^{-1/2}.
$
As a vector space ${\cal F}_{-1/2}$ is isomorphic to $C^{\infty}(S^1)$
and the $\Vect(S^1)$-action on ${\cal F}_{-1/2}$ is given by the Lie derivative:
$$
L^{-1/2}_X(\phi)=(X\phi'-(1/2)X'\phi)(dx)^{-1/2}.
$$

\noindent 
There exists a natural Lie superalgebra structure on the space
$
\Vect(S^1)\oplus{\cal F}_{-1/2},
$
The anticommutator 
$$
[\;,\;]_+:{\cal F}_{-1/2}\otimes{\cal F}_{-1/2}\longrightarrow\Vect(S^1)
$$
is just the product of tensor-densities:
$$
[\xi(x)(dx)^{-1/2},\eta(x)(dx)^{-1/2}]_+:=
\xi(x)\eta(x)\frac{d}{dx}.
$$

\noindent 
Thus, the commutator in the defined Lie superalgebra is given by the following formula:
$$
\bigg[(X,\xi),(Y,\eta)\bigg]=
([X,Y]_{{\rm Vect}(S^1)}+\xi\cdot\eta,\;L_X(\eta)-L_Y(\xi))
$$

\vskip 0,3cm

\noindent 
{\bf Definition 3.1}. There exists a unique (modulo isomorphism) nontrivial central
extension of the defined Lie superalgebra. It can be given by the following 2-cocycle:
$$
\Omega\bigg((Xd/dx,\xi(dx)^{-1/2}),(Yd/dx,\eta(dx)^{-1/2})\bigg)=
\int_{0}^{2\pi}(X''Y'+2\xi'\eta')dx
\eqno (17)
$$

\vskip 0,3cm

\noindent The Lie superalgebra defined by this central extension is called the 
{\it Ramond} algebra.

\vskip 0,3cm

\noindent {\bf Remark}. The even part of the Ramond Lie superalgebra 
coincides with the Virasoro algebra.

\vskip 0,3cm

\noindent Consider now the space of {\it anti-periodic} $-1/2$-densities on 
$S^1$:
$$
\xi(x)(dx)^{-1/2},\;\;\;\;\;\xi(x+2\pi)=-\xi(x).
$$
This space is also a $\Vect(S^1)$-module. 
Let us denote it: ${\cal F}^{(-)}_{-1/2}$.
Note that the product of two anti-periodic 
$-1/2$-densities is a (``periodic'') vector field well-defined on $S^1$.

\vskip 0,3cm

\noindent {\bf Definition 3.2}. The same formulae as above define a 
Lie superalgebra structure on the space 
$$
\Vect(S^1)\oplus{\bf R}\oplus{\cal F}^{(-)}_{-1/2}.
$$
This Lie superalgebra is called the {\it Neveu-Schwarz}
algebra.

\vskip 0,3cm

\noindent {\bf Remarks}. (a) The Lie superalgebras on
$\Vect(S^1)\oplus{\cal F}_{-1/2}$
and $\Vect(S^1)\oplus{\cal F}^{(-)}_{-1/2}$
can be defined as the Lie superalgebras of contact vector fields
on $S^{1|1}$ and ${\bf RP}^{1|1}$ reciprocally
(see [21]).

(b) The Ramond and Neveu-Schwarz superalgebras are particular cases
of a series of so-called string superalgebras
(see [21]).

\bigskip
\noindent
{\large 3.3 Coadjoint representation}
\medskip
\noindent
The (regularized) dual space to the Ramond algebra is
naturally isomorphic to:
$$
{\cal F}_2\oplus{\bf R}\oplus{\cal F}_{3/2}.
$$
Indeed, the module ${\cal F}_{3/2}$ is dual to ${\cal F}_{-1/2}$
with respect to the pairing
$$
\langle\phi(x)(dx)^{3/2},\xi(x)(dx)^{-1/2}\rangle=
\int_{0}^{2\pi}\phi(x)\xi(x)dx.
$$
Thus, the regular dual space to the Ramond algebra
consists of the elements:
$
(u,c,\phi)=(u(x)(dx)^2,c,\phi(x)(dx)^{3/2}).
$

In the same way, the regular dual space to the Neveu-Schwarz algebra
is:
$$
{\cal F}_2\oplus{\bf R}\oplus{\cal F}^{(-)}_{3/2},
$$
where ${\cal F}^{(-)}_{3/2}$ is the space of antiperiodic $3/2$-densities.

\bigskip
\noindent
{\bf Lemma 3.3.} {\sl The coadjoint representation of the Ramond 
and Neveu-Schwarz superalgebras
are given by the formula:
$$
ad^*_{(X,\xi)}\left(
\matrix{
u(dx)^2\hfill\cr
c\hfill\cr
\phi(dx)^{3/2}\hfill\cr
}
\right)=
\left(
\matrix{
(Xu'+2X'u-c\cdot X'''\;+\xi\phi'/2+3\xi'\phi/2)(dx)^2\hfill\cr
0\hfill\cr
(X\phi'+3X'/2\;+u\xi-2c\cdot\xi'')(dx)^{3/2}\hfill\cr
}
\right)
\eqno(18)
$$   }

\noindent {\bf Proof}. The formula (18) 
can be obtained directly from the definition of the coadjoint 
representation. The easy calculations are similar to those
from the proof of Lemma 1.3.

\vskip 0,3cm

\noindent 
The Sturm-Liouville operator appears as the coadjoint action of 
the odd part of the Ramond 
and Neveu-Schwarz superalgebras. Indeed,
$$
ad^*_{(0,\xi)}(u,c,0)=\bigg(-2c\frac{d^2}{dx^2}+u\bigg)\xi.
$$

\bigskip
\goodbreak
\noindent
\item{\large 3.4} {\Large Projective equivariance and Lie 
superalgebra $osp(1|2)$}
\medskip
\noindent
Consider the Lie superalgebra generated by the vector fields 
(11)
and two more odd generators:
$$
\sin\left(\frac{x}{2}\right)(dx)^{-1/2}
\;\;\;\hbox{and}\;\;\;
\cos\left(\frac{x}{2}\right)(dx)^{-1/2}
$$
This Lie superalgebra is a subalgebra of the Neveu-Schwarz superalgebra
isomorphic to the $osp(1|2)$, that has a natural interpretation
as the algebra of symmetries of the projective superspace 
${\bf P}^{1|1}$.

As in the case of the Virasoro algebra (cf. Section 2)
it is possible to write the cocycle giving the central extension
in a canonic ($osp(1|2)$-invariant) form.
\medskip
\noindent
{\bf Lemma 3.4.} {\sl The 2-cocycle 
$$
\bar\Omega\bigg((Xd/dx,\xi(dx)^{-1/2}),(Yd/dx,\eta(dx)^{-1/2})\bigg)=
\int_{0}^{2\pi}((X'''+X')Y+2(\xi''+4\xi)\eta)dx
$$
is the unique (up to a constant) nontrivial 2-cocycle on
the Lie superalgebra $\Vect(S^1)\oplus{\cal F}^{(-)}_{-1/2}$
equivariant with respect to the subalgebra $osp(1|2)$.}
\par

\noindent {\bf Proof.} Similar to those of the proof of 
Proposition 2.1.

\bigskip
\noindent  
\item{\Large 4} {\Large Invariants of coadjoint representation
of the Virasoro group}
\bigskip
\noindent
It follows from Theorem 1 that the
invariants of the coadjoint representation of the Virasoro group
are the invariants of the 
$\Diff^+(S^1)$-action on the space of
Sturm-Liouville operator (1).
This is quite old and classical problem was considered in
[18],[19] and in [13],[27], [30], [10]
in the context of the Virasoro algebra.

In this section we will describe the invariants of the
$\Diff^+(S^1)$-action following [13] and [27].

\bigskip
\noindent
{\large 4.1 Monodromy operator as a conjugation class of $\widetilde{SL}(2,{\bf R})$}
\medskip
\noindent
Consider the Sturm-Liouville equation $2c\psi''+u(x)\psi=0$.
Since the potential $u(x)$ is a periodic function, the translation
$$
M\psi(x)=\psi(x+2\pi)
$$
defines is a linear operator on the space of solutions.
This operator is called the {\it monodromy operator}.

\vskip 0,3cm

\noindent We need the following two remarks.

\vskip 0,3cm

\noindent (1) The monodromy operator defines a conjugation class 
of the group $SL(2,{\bf R})$.
Indeed, the Wronsky determinant of any two solutions
$$
W(\psi_1,\psi_2)=\psi_1\psi_2'-\psi_1'\psi_2
$$ 
is a constant function.
Thus, $W$ defines a bilinear skew-symmetric form on the space of solutions
and operator $M$
preserves $W$. Now, an arbitrary choice of the basis
$\psi_1,\psi_2$ such that $W(\psi_1,\psi_2)=1$ associates to $M$
a matrix from $SL(2,{\bf R})$. The conjugation class of this matrix does not depend
on the choice of the basis.

\vskip 0,3cm

\noindent (2) Moreover, the monodromy operator defines a conjugation 
class of the
universal covering $\widetilde{SL}(2,{\bf R})$.
Indeed,
for every value $x=x_0$, identify the space of solutions
with ${\bf R}^2$ choosing the initial conditions:
$T_{x_0}:\psi\mapsto(\psi(x_0),\psi'(x_0))$.
Define a family of linear operators on the space of solutions:
$$
T(x):=T^{-1}_{x}\circ T_{0}
$$
The family $T(x)$ joins the monodromy operator:
$M=T(2\pi)$ with the identity: $T(0)=\hbox{Id}$.
It can be lift (up to a conjugation) to $\widetilde{SL}(2,{\bf R})$.

\vskip 0,3cm

\noindent We will confound the monodromy operator with the corresponding
conjugation class of $\widetilde{SL}(2,{\bf R})$.

\bigskip
\noindent
{\large 4.2 Classification theorem}
\medskip
\noindent
The following theorem is the classification of the invariants of
the $\Diff^+(S^1)$-action on the space of Sturm-Liouville operators.
According to Theorem 1, it classifies also the invariants of the coadjoint
action of the Virasoro group.
Various approaches to the classification see in
[18],[19],[13] and [27].

\medskip
\noindent
{\bf Theorem 2.} {\sl The monodromy operator
is the unique invariant of the $Diff^+(S^1)$-action.}

\noindent {\bf Proof}. Let us give a simple proof (different from 
those of [18],[19],[13] and [27]) based on [23].

First, it is clear that the monodromy operator
{\it is} an invariant,
since the 
$\Diff^+(S^1)$-action is just a coordinate transformation
and the monodromy operator is defined intrinsically.

To prove that there is no more (independent of $M$) invariants, we will
use the homotopy method. One should show that:

(a) Every two Sturm-Liouville operators 
with the same monodromy are homotopic to each other
in the class of operators with the fixed monodromy.
In other words, the set of operators with fixed monodromy is connected.

(b) Given a smooth family of operators with fixed monodromy:
$$
L_s=2c\frac{d^2}{dx^2}+u_s(x),\;\;\;s\in[0,1],
$$
there exists a vector field $X\in\Vect(S^1)$ such that
$$
Xu'+2X'u+cX'''=\dot u,\;\;\;
\hbox{where}\;\;\;
\dot u=\left.\frac{\partial}{\partial s}u_s(x)\right|_{s=0}
\eqno (19)
$$
\vskip 0,3cm

\noindent 
Statements (a) and (b) imply that every two Sturm-Liouville operators
$L_1$ and $L_1$
with the same monodromy are on the same $\Diff^+(S^1)$-orbit.
Indeed, there exists a family $L_s$ of operators with the fixed monodromy
and a family $X_s\Vect(S^1)$ of solutions of the homotopy equation 
(19)
for each $s$. Now, there exists a flow corresponding to this family
($S^1$ is compact!), this is a diffeomorphism which maps $L_1$ to $L_2$.

\vskip 0,3cm

\noindent Let us first prove (a).
To each Sturm-Liouville operator, associate a family of mapping
$T(x)$ (see Section 4.1). If $L_1,L_2$ are two operators with the
same monodromy, then, the corresponding families are homotopic 
(as two curves in $PSL_2({\bf R}))$.
Fixe this homotopy: $T(x)^{\tau}$.
Now, to define a family of Sturm-Liouville operators $L_{\tau}$
joining $L_1$ and $L_2$,  one associates a Sturm-Liouville
oparator 
to each family $T(x)^{\tau}\in PSL_2({\bf R})$, for fixed $\tau$
(this is a standard procedure).

\vskip 0,3cm

\noindent Let us now prove (b).
\medskip
\noindent
{\bf Lemma 4.1.} {\sl Given a basis $\psi_1^s,\psi_2^s$ of solutions
of the equation $L_s\psi=0$ such that $W(\psi_1^s,\psi_2^s)\equiv1$,
the vector field
$$
X=\frac{1}{2}\left|
\matrix{
\dot \psi_1&\dot \psi_2\hfill\cr
\psi_1&\psi_2\hfill\cr
}
\right|
\eqno (20)
$$
is a solution of the equation (19).} \par

\noindent {\bf Proof}. Taking the derivative of the equality 
$L_s\psi^s=0$, one gets:
$$
\dot u\psi^{s=0}+L_{s=0}\dot\psi=0.
$$
To solve the equation (19), it is sufficient to 
find a vector fields $X$ such that the Lie derivative
$L_X(\psi)=\dot\psi$. 
Indeed, it follows from the definition (9) of the $\Vect(S^1)$-action
on the space of Sturm-Liouville operators.

Let us look for a vector field $X$ such that
$L_X(\psi_1^{s=0})=\dot\psi_1$ and
$L_X(\psi_2^{s=0})=\dot\psi_2$. This gives a system of linear equations:
$$
\left\{
\matrix{
X\psi'_1-(1/2)X'\psi_1=\dot\psi_1\cr
X\psi'_2-(1/2)X'\psi_2=\dot\psi_2 \, . \cr
}
\right.
$$
Taking $X$ and $X'$ as independent arguments,
one obtains formally:
$$
X=
\frac{1}{2}
\left|
\matrix{
\dot \psi_1&\dot \psi_2\hfill\cr
\psi_1&\psi_2 \hfill\cr
}
\right|,\;\;\;\;
X'=
\left|
\matrix{
\dot \psi_1&\dot \psi_2\hfill\cr
\psi_1&\psi_2\hfill\cr
}
\right| \, .
$$
Now, let us verify that $X'=dX/dx$. Indeed,
$$
\frac{dX}{dx}=\frac{1}{2}
\left|
\matrix{
\dot \psi_1'&\dot \psi_2'\hfill\cr
\psi_1&\psi_2\hfill\cr
}
\right|+
\frac{1}{2}
\left|
\matrix{
\dot \psi_1&\dot \psi_2\hfill\cr
\psi_1'&\psi_2'\hfill\cr
}
\right| \, .
$$
The two terms in the right hand side coincide since:
$$
\frac{d}{ds}
\left|
\matrix{
\psi_1'&\psi_2'\hfill\cr
\psi_1&\psi_2\hfill\cr
}
\right|=
\left|
\matrix{
\dot \psi_1'&\dot \psi_2'\hfill\cr
\psi_1&\psi_2\hfill\cr
}
\right|+
\left|
\matrix{
\psi_1'&\psi_2'\hfill\cr
\dot \psi_1&\dot \psi_2\hfill\cr
}
\right|=0,
$$
and therefore $X'=dX/dx$.

\vskip 0,3cm

\noindent Lemma 4.1 is proven.

\vskip 0,3cm

\noindent Let us show that
Theorem 2 follows from the lemma.
Indeed, since the monodromy operator does not depend on $s$, 
the basis $\psi_1^s,\psi_2^s$ can be chosen in such a way that the
corresponding monodromy
matrix does not depend on $s$. Then, the solution (20)
is periodic: $X(x+2\pi)=\det(M)X(x)=X(x)$.

\vskip 0,3cm

\noindent Theorem 2 is proven.

\vskip 0,3cm

\noindent {\bf Remark}. Recall that the solutions of a Sturm-Liouville
equation have a sense of $-1/2$-tensor-densities.
Therefore, the quadratic expression (20) is indeed a vector field.

\vskip 0,3cm

\noindent The K\"ahler geometry of the coadjoint orbits of the 
Virasoro group
has been studied in A.A. Kirillov's works [15],[16].

\bigskip
\noindent
\item{\Large 5}{\Large Extension of the Lie algebra 
of first order linear differential operators on $S^1$
and matrix analogue of the Sturm-Liouville
operator}
\medskip
\noindent
This section follows the recent work
[22]. 
We will show that the Kirillov method
is valid in a more general framework then the Virasoro algebra.

\bigskip
\noindent
{\large 5.1 Lie algebra 
of first order differential operators on $S^1$ and its central 
extensions}
\medskip
\noindent
Consider the Lie algebra of first order linear differential operators on $S^1$:
$$
A=X(x)\frac{d}{dx}+a(x)
\eqno (21)
$$
(This Lie algebra is in fact the semi-direct product
of $\Vect(S^1)$ by the module of functions ${\cal F}_0$).

This Lie algebra has three nonisomorphic
central extensions (cf. [25]).
The first one is given by the Gelfand-Fuchs cocycle and
two more extensions are given by the non-trivial $2$-cocycles: 
$$
\matrix{
\omega'((X\frac{d}{dx},\;a),\;(Y\frac{d}{dx},\;b))=
\displaystyle\int_{S^1}(X''(x)b(x)-Y''(x)a(x))dx\hfill\cr\noalign{\bigskip}
\omega''((X\frac{d}{dx},\;a),\;(Y\frac{d}{dx},\;b))=
\displaystyle2\int_{S^1}a(x)b'(x)dx\hfill\cr\noalign{\bigskip}
}
\eqno (22)
$$

\vskip 0,3cm

\noindent {\bf Definition 5.1}.
Let us denote $\cal G$ the Lie algebra defined
on the space  $Vect (S^1)\oplus C^{\infty}(S^1)\oplus{\bf R}^3$
as the {\it universal central extension} of the Lie algebra of the 
operators (21).
This means, $\cal G$ is the Lie algebra defined by the commutator:
$$
\bigg[(X\frac{d}{dx},a,{\bf \alpha}),\;
(Y\frac{d}{dx},b,{\bf \beta})\bigg]=
\bigg( (XY'-X'Y)\frac{d}{dx} ,\;Xb'-Ya',\; {\bf \omega}\bigg)
$$
where ${\bf \alpha}=(\alpha_1,\alpha_2,\alpha_3),\;
{\beta}=(\beta_1,\beta_2,\beta_3)\in {\bf R}^3$ are in the center and
$$
\matrix{\omega=
(\omega((X\frac{d}{dx},\;a),\;(Y\frac{d}{dx},\;b)),\;\;
 \omega'((X\frac{d}{dx},\;a),\;(Y\frac{d}{dx},\;b)),\;\;
\omega''((X\frac{d}{dx},\;a),\;(Y\frac{d}{dx},\;b))).}
$$

\vskip 0,3cm

\bigskip
\noindent
{\large 5.2 Matrix Sturm-Liouville operators}
\medskip
\noindent
The space of
matrix linear differential operators
on $C^{\infty}(S^1)\oplus C^{\infty}(S^1)$:
$$
{\cal L}=\left(
\matrix{
-2c_1\displaystyle\frac{d^2}{dx^2} +u(x) & 2c_2\displaystyle\frac{d}{dx}+v(x) \hfill\cr\noalign{\smallskip}
-2c_2\displaystyle\frac{d}{dx}+v(x) & 4c_3\hfill\cr\noalign{\smallskip}
}
\right)
\eqno (23)
$$
where $c_1,c_2,c_3\in {\bf R}$ and $u=u(x),v=v(x)$
are $2\pi$-periodic functions
was defined in [22].
It was shown that this space gives a geometric realization for the dual
space of the Lie algebra $\cal G$.

\vskip 0,3cm

\noindent 
The $Vect (S^1)$-action on the space of operators (23) is defined,
as in the case of Sturm-Liouville operators (1),
by commutator with the Lie derivative.
We consider 
 ${\cal L}$ as an operator on $Vect (S^1)$-modules:
$$
{\cal L}:{\cal F}_{-\frac{1}{2}}\oplus{\cal F}_{\frac{1}{2}}\to
{\cal F}_{\frac{3}{2}}\oplus{\cal F}_{\frac{1}{2}}.
$$

\vskip 0,3cm

\noindent {\bf Remark}. The choice of degrees of tensor-densities 
in this formula is the {\it unique} choice such that the operators (23)
are {\it selfadjoint}.

\vskip 0,3cm

\noindent There exists 
a natural action of the 
Lie algebra of first order differential operators (21)
 the space of operators (23).

\bigskip
\noindent
{\large 5.3 Action of Lie algebra of differential operators}
\medskip
\noindent
There exists a nice family of modules over the Lie algebra of
operators (21). Consider the space
$$
{\cal F}_{\lambda}\oplus{\cal F}_{\lambda+1}
$$
It is defined by the formula:
$$
T^{(\lambda)}_{\displaystyle(X(x)\frac{d}{dx}+a(x))}
\left(
\matrix{
\phi(x)\hfill\cr\noalign{\smallskip}
\psi(x)\hfill\cr\noalign{\smallskip}
}
\right)=
\left(
\matrix{
L_{X\frac{d}{dx}}^{(\lambda)}\;\phi(x)\hfill\cr\noalign{\smallskip}
L_{X\frac{d}{dx}}^{(\lambda+1)}\;\psi(x) -{\lambda}a'(x){\phi}(x)\hfill\cr\noalign{\smallskip}
}
\right)
\eqno (24)
$$
\vskip 0,3cm

\noindent The action on the space of operators (23) is defined in analogous way as
the action of $\Vect(S^1)$ on the space of Sturm-Liouville operators
given by the formula (9).
Put:
$$
\Bigl[T_{\displaystyle(X\frac{d}{dx}+a)},{\cal L}\Bigr]:=
T^{(1/2)}_{\displaystyle(X\frac{d}{dx}+a)}\;\circ\;{\cal L}-
{\cal L}\;\circ\;T^{(-1/2)}_{\displaystyle(X\frac{d}{dx}+a)}
\eqno (25)
$$
{\bf Theorem 5.2 (see [22]).} {\sl The action 
(25) coincides with the coadjoint action of the 
Lie algebra of first order linear differential operators.}
\smallskip

\noindent {\bf Proof}. The explicit formula for the action (25) is:
$$
\Bigl[ T_{\displaystyle(X\frac{d}{dx}+a)},{\cal L}\Bigr]=\left(
\matrix{
\matrix{
Xu'+2X'u-c_1X'''\hfill  \cr
+va'+c_2a''\hfill  \cr} & 
\matrix{
Xv'+X'v-c_2X''\hfill \cr
+2c_3a'\hfill \cr}\hfill\cr\noalign{\bigskip}
\matrix{
Xv'+X'v-c_2X'' \hfill  \cr
+2c_3a'\hfill  \cr} & 0\hfill\cr\noalign{\bigskip}
}\right)
$$
One easily verifies that this is precisely the coadjoint action
of the Lie algebra of differential operators (23)
(see [22] for the details).

\bigskip
\noindent
{\large 5.4 Generalized Neveu-Schwarz superalgebra}
\medskip
\noindent
The space of matrix analogues of the Sturm-Liouville operators
(23) was found in [22] using a Lie superalgebra
generalizing the Neveu-Schwarz algebra.

\vskip 0,3cm

\noindent {\bf Definition 5.3}.
 Consider the ${\bf Z}_2$-graded vector space : 
 ${\cal S}={\cal S}_0\oplus{\cal S}_1$.
where 
${\cal S}_0=\cal G$
the extension of the Lie algebra of operators (23).
 and ${\cal S}_1$ the $\cal G$-module:
$$
{\cal S}_1={\cal F}_{-\frac{1}{2}}\oplus{\cal F}_{\frac{1}{2}}.
$$
The even part ${\cal S}_0$ acts on ${\cal S}_1$ according to 
(24).
Let us define the 
{\it anticommutator}
$[\;,\;]_+:{\cal S}_1\otimes{\cal S}_1\to{\cal S}_0$:
$$
\Bigl[(\phi,\;\alpha),\;
(\psi,\;\beta)\Bigr]_+=
(\phi\psi\displaystyle\frac{d}{dx},\;\phi\beta+\alpha\psi,\;
{\bf \sigma_+})
$$
where ${\bf \Omega_+}=(\Omega,\Omega',\Omega'')$,
where $\Omega$ is the Ramond -- Neveu-Schwarz cocycle
(17) and $\Omega',\Omega'')$
are the continuations of the cocycles (22):
$$
\matrix{
\Omega'((\phi,\alpha),(\psi,\beta))=
-2\displaystyle\int_{S^1}(\phi'(x)\beta(x)+\alpha(x)\psi'(x))dx\hfill\cr\noalign{\bigskip}
\Omega''((\phi,\alpha),(\psi,\beta))=
4\displaystyle\int_{S^1}\alpha(x)\beta(x)dx\hfill\cr\noalign{\bigskip}
}
$$

\vskip 0,3cm

\noindent
{\bf Theorem 5.4 (see [22]).}  {\sl ${\cal S}$ is a Lie superalgebra.}
\medskip

\noindent 
The differential operators (23) can be defined as a part of the coadjoint
action of the superalgebra $\cal G$. 
Namely, one obtains:
$$
ad^*_{\left(\matrix{
0\hfill\cr
0\hfill\cr
\phi(dx)^{-\frac{1}{2}}\hfill\cr
\alpha(dx)^{\frac{1}{2}}\hfill\cr
}
\right)}
\left(
\matrix{
u\hfill\cr
v\hfill\cr
{\bf c}\hfill\cr
0\hfill\cr
0\hfill\cr
}\right)
=\left(
\matrix{
0\hfill\cr
0\hfill\cr
0\hfill\cr
-2c_1\phi''+u\phi+v\alpha+2c_2\alpha'\hfill\cr
-2c_2\phi'+v\phi+4c_3\alpha \hfill\cr
} 
\right)
$$

\vskip 0,3cm

\noindent {\bf Remark}. The Lie algebra $\cal G$ considered 
in this section, is just an example from the 
series of seven Lie algebras generalizing the Virasoro algebra 
(see [25]).
It turns out that Kirillov's method works for also for the other Lie algebra from
this series (work in preparation of P. Marcel).
It would be very interesting to apply this method to other
Virasoro type Lie algebras (see [26]).

\bigskip
\noindent
\item{\Large 6} {\Large Geometrical definition of the Gelfand-Dickey bracket and the relation
to the Moyal-Weil star-product}
\bigskip
\noindent
In this section we follow [24].
We consider another generalization of the Virasoro algebra:
the so-called {\it second Adler-Gelfand-Dickey} 
Poisson structure, which is also known as the classical $W$-algebras
in the physics literature.
The Adler-Gelfand-Dickey bracket is (an infinite-dimen\-sional)
Poisson bracket on the space of $n$-th order differential operators on $S^1$.
(We will consider here only the first nontrivial case corresponding to
the space of third order linear differential operators)

We will show that the Gelfand-Dickey bracket is related to the
well-known {\it Moyal-Weyl star-product}. 

The main idea is to consider arguments of differential operators as tensor-densities
and use the $PSL_2$-equivariance of all the operations.

\bigskip
\noindent
{\large 6.1 Moyal-Weyl star-product}
\medskip
\noindent
Consider the standard symplectic plane $({\bf R}^2,dp\wedge dq)$,
where $p,g$ are linear coordinates.
The space of functions on  ${\bf R}^2$ is a Lie algebra with
respect to the Poisson bracket:
$$
\{F,G\}=F_pG_q-F_qG_p,
$$
where $F_p=\partial F/\partial p$.

\vskip 0,3cm

\noindent The following operation:
$$
F\star_{\hbar}G=FG+\frac{\hbar}{2}\{F,G\}+\cdots+
\frac{\hbar^m}{2^mm!}\{F,G\}_m+\cdots
$$
where 
$$
\{F,G\}_m=
\sum_{i=0}^{m}(-1)^i
\left(
\matrix{
m\cr
i\cr
}
\right)
\frac{\partial ^mF}{\partial p^{m-i}\partial q^i}
\frac{\partial ^mG}{\partial p^i\partial q^{m-i}}
$$
is called the Moyal-Weyl star-product on ${\bf R}^2$.
Here $\hbar$ is a formal parameter
and the operation $\star_{\hbar}$ is with values in formal series in $\hbar$. 
(In the case polynomials, one can consider $\hbar$ as a number).
The operation $\star_{\hbar}$ is associative.

\vskip 0,3cm

\noindent 
The Moyal-Weyl star-product is a very popular object in deformation quantization.

\bigskip
\noindent
{\large 6.2 Moyal-Weyl star-product on tensor-densities,
the transvectants}
\medskip
\noindent
{\bf Isomorphism 6.1.} {\sl There exists a natural isomorphism
between the space ${\cal F}_{\lambda}$ (of tensor-densities 
of degree $\lambda$ on $S^1$) and the space of functions
on ${\bf R}^2\setminus\{0\}$ homogeneous of degree $-2\lambda$.
For the affine parameter on $S^1$: $t=\hbox{tg}(x)$ this isomorphism
is given by the formula:
$$
\phi(t)(dt)^{\lambda}\;\longmapsto\;
p^{-2\lambda}\phi(\frac{q}{p})
\eqno (26)
$$  }

\noindent Indeed, a function corresponding to a vector field $X$
is: $p^2X(q/p)$. Verify, that the Lie derivative corresponds to the
Poisson bracket.

\vskip 0,3cm

\noindent The isomorphism (26) lifts the Moyal-Weyl star-product 
to the space of tensor-densities.
\medskip
\noindent
{\bf Lemma 6.2.} {\sl The terms of this star-product are as follows:
$$
\{\phi, \psi \}_m
= 
\frac{m!}{2^m}\sum_{i+j=m} (-1)^i m!
{2\lambda+m-1 \choose i} {2\mu+m-1 \choose j}  
{\phi}^{(i)} {\psi}^{(j)} 
\eqno (27)
$$
where $\phi\in{\cal F}_{\lambda},\psi\in{\cal F}_{\mu}$,
${\phi}^{(i)}=d^i\phi/dx^i$
and 
$$
{k\choose i}=k(k-1)\cdots(k-i+1).
$$  }

\noindent {\bf Proof.} Straightforward.

\vskip 0,3cm

\noindent It turns out that the operations (27)
coincides (up to the constant $m!/2^m$) with so-called Gordan's {\it transvectants}.
This operations can be defined as bilinear maps
$$
{\cal F}_{\lambda}\otimes{\cal F}_{\mu}\;\to\;
{\cal F}_{\lambda+\mu+m}
$$
equivariant with respect to the action of the Lie algebra $sl_2({\bf R}$
defined by (11) (projectively invariant).

\vskip 0,3cm

\noindent {\bf Remark}. The isomorphism (26) is, in fact, given
by the standard projective structure on $S^1$. Indeed, $t$ is
the corresponding projective parameter.
Given an arbitrary projective structure on $S^1$,
one defines an isomorphism (analogue of (26)) between tensor-densities on $S^1$
and homogeneous functions on ${\bf R}^2$.

\bigskip
\noindent
{\large 6.3 Space of third order linear differential operators as a
$\Diff^+(S^1)$-module}
\medskip
\noindent
Consider the space of third order linear differential operators
$$
A=\frac{d^3}{dx^3}+u(x)\frac{d}{dx}+v(x)
\eqno (28)
$$
This space plays the same role that the space of Sturm-Liouville
operators (1) in the case of Virasoro algebra.
However, the Adler-Gelfand-Dickey bracket is not a Lie-Poisson structure.
We refer [1] and [7] for the original definition and
[4] for another one related to the Kac-Moody algebras.

\vskip 0,3cm

\noindent The subject of this section was known 
to the classics (see [29],[3]) ... and was forgotten by the
contemporary experts.

\vskip 0,3cm

\noindent {\bf Definition 6.3}. The $\Diff^+(S^1)$-action on the 
space of operators (28) is defined by the formula:
$$
g^*(A):=g^*_{2}\circ A\circ(g^*_{-1})^{-1}
$$

\vskip 0,3cm

\noindent This means that the operator $A$ is considered as acting 
from the space of vector fields on $S^1$ with values
in the space of quadratic differentials:
$$
A:{\cal F}_{-1}\to{\cal F}_2.
$$
The corresponding action of $X(x)d/dx\in\Vect(S^1)$ is:
$$
\ad L_X(A):=L^{(2)}_X\circ A-A\circ L^{(-1)}_X
$$

\noindent Let us give the explicit formulae for 
$\Diff^+(S^1)$- and $\Vect(S^1)$-action.
It is convenient to decompose the operator (28)
as a sum of its skew-symmetric and symmetric parts:
$$
A=\frac{d^3}{dx^3}+u(x)\frac{d}{dx}+\frac{u(x)}{2}+w(x),
$$
where $w(x)=v(x)-u(x)/2$.
\smallskip
\noindent
{\bf  Proposition 6.4 (see [29],[3]).} {\sl A diffeomorphism
$f$ transform an operator $A$ into an operator of the form (28)
with coefficients:
$$
\matrix{
u^f=u\circ f(f')^2+2S(f)\hfill\cr\noalign{\smallskip}
w^f=w\circ f(f')^3.\hfill\cr
}
$$  }
\smallskip

\noindent This means, $u$ transforms as a potential of the 
Sturm-Liouville operator:
$
4\frac{d^2}{dx^2}+u(x)
$
and $w$ has the sense of {\it cubic differential}: $w=w(x)(dx)^3$.
\smallskip
\noindent
{\bf Corollary.} {\sl The projection from the space of third order
operators (28) to the space of Sturm-Liouville operators:
$$
\frac{d^3}{dx^3}+u(x)\frac{d}{dx}+v(x)\;\longmapsto\;
4\frac{d^2}{dx^2}+u(x)
$$
is $\Diff(S^1)$-equivariant (does not depend on the choice of the parameter
$x$).}\par

The corresponding action of a vector field $X(x)d/dx\in\Vect(S^1)$
associates to $A$ a first order operator:
$
\ad L_X(A)=u^X\frac{d}{dx}+\frac{u^X}{2}+w^X,
$
where
$$
\matrix{
u^X=Xu'+2X'u+2X'''\hfill\cr\noalign{\smallskip}
w^X=Xw'+3X'w.\hfill\cr
}
$$  

\vskip 0,3cm

\noindent {\bf Remark}. The geometric interpretation of $u$ and $w$ is related
to the projective projective differential geometry of plane curves
(associated to differential operators (28)).
Namely, $u$ is interpreted as the projective curvature and $w$ leads to
the notion of projective length element: $ds=(w)^{1/3}$
(see [29],[3] and also [11]).

\bigskip
\goodbreak
\noindent
{\large 6.4 Second order Lie derivative}
\medskip
\noindent
The notion of {\it second order Lie derivative} considered below was
introduced in [24] (see [5] for a general definition
in the multi-dimensional case).
The question is as follows: given a second order contravariant tensor
field $X\in{\cal F}_2$:
$$
Z=Z(x)(dx)^{-2},
$$
is it possible to define an ``action'' of $Z$ on geometric quantities
(like tensor-densities etc.) analogous to the Lie derivative along
a vector field?

The answer is negative. There is no $\Diff(S^1)$-equivariant
bilinear differential operators
$$
{\cal F}_2\otimes{\cal F}_{\lambda}\to{\cal F}_{\lambda},
$$
for general values of $\lambda$ (cf. [12]) 
and so, one can not define such an action
intrinsically.

To define the second order Lie derivative, we fix a 
projective structure on $S^1$.
\smallskip

\noindent {\bf Definition 6.5.}
The second order Lie derivative over contravariant tensor field
of degree $2$: $Z=Z(x)(dx)^{-2}$ is a linear map
$$
L^2_Z:{\cal F}_{\lambda}\to{\cal F}_{\lambda}
$$
given by:
$$
L^2_Z(\phi):=\{Z,\phi\}_2
$$
\par

\noindent {\bf Remark}. Note, that the operations $\{\;,\;\}_m,\;m\geq2$ are defined
if one fix a projective structure (cf. Section 6.2).

\bigskip
\noindent
{\large 6.5 Adler-Gelfand-Dickey Poisson structure}
\medskip
\noindent
A Poisson structure on a manifold
is given by a linear map on each cotangent space with values
in the tangent space (satisfying the Jacobi condition).
Thus, to define a Poisson structure on a vector space, it is
sufficient to associate a vector field
to every linear functional.

Every linear functional on the space of operators (28) is a linear combination of:
$$
\langle l^1_X,A\rangle=\int X(x)u(x)dx,\;\;\;
\langle l^2_Z,A\rangle=\int Z(x)w(x)dx
$$
where $X=X(x)d/dx,Z=Z(x)(dx)^{-2}$.

\vskip 0,3cm

\noindent {\bf Definition 6.6}. The Adler-Gelfand-Dickey Poisson structure
on the space of operators (28) associates to 
a linear functionals
vector fields given by the commutator with the Lie derivative:
$$
\matrix{
\dot A_X:=[L_X,A]\hfill\cr\noalign{\smallskip}
\dot A_Z:=[L^2_Z,A]\hfill\cr
}
$$
(see [24] for the details).

\vskip 0,3cm

\noindent The Adler-Gelfand-Dickey Poisson structure
is a very interesting and popular object in Mathematical Physics.
This way to define it seems to be natural in the spirit of Section 1.
\vskip 0,3cm
\noindent {\bf Addendum.} Recently Kirillov's method has been applied 
for a new class of infinite-dimensional Lie algebras, see [31, 32].

\vskip 1.8cm
\goodbreak

{\Large References}
\bigskip
\noindent
\item{[1]} M. Adler, {\it On a trace functional for formal 
pseudo-differential operators and the symplectic structure 
of the Korteweg-de Vries type equation}, Invent. Math.
50:3 (1987) 219-248.

\item{[2]} R. Bott, {\it On the characteristic classes of groups of diffeomorphisms},
Enseign. Math. 23:3-4 (1977), 209-220.

\item{[3]} E. Cartan, {\it Le\c cons sur la th\'eorie des espaces
\`a connexion projective}, Gauthier -Villars, Paris, 1937.

\item{[4]} V.G. Drinfel'd \& V.V. Sokolov, {\it Lie algebras
and equations of Korteweg - De Vries type}, J. Soviet Math.
30 (1985), 1975 - 2036.

\item{[5]} C. Duval \& V. Ovsienko,
{\it Space of second order linear differential operators as a module over the 
Lie algebra of vector fields},  Advances in Math. 132 (1997), no.2, 316-333. 

\item{[6]} H. Gargoubi \& V. Ovsienko,
{\it Space of linear differential operators on the real line as a module over the 
Lie algebra of vector fields},  IMRN (1996), N.5, 235-251.

\item{[7]} I.M. Gel'fand \& L.A. Dikii - {\it A family of
hamiltonian structures connected with integrable nonlinear differential
equations} in: I.M. Gel'fand collected papers (S.G. Gindikin et al,
eds), Vol. {\bf 1}, Springer, 1987, 625-646.

\item{[8]} I.M. Gel'fand \& D.B. Fuchs, {\it Cohomology of the Lie algebra of vector
fields on the circle}, Funct. Anal. Appl. 2:4 (1968), 342-343.

\item{[9]} P. Gordan, Invariantentheorie, Teubner, Leipzig, 1887.

\item{[10]} L. Guieu, {\it Nombre de rotation, structures g\'eom\'etriques
sur un cercle et groupe de Bott-Virasoro}, Ann. Inst. Fourier 46
(1996), no.4, 971-1009.

\item{[11]} L. Guieu \& V. Ovsienko, 
{\it Structures symplectiques sur les espaces de courbes projectives et
affines}, J. Geom. Phys. 16 (1995) 120-148.

\item{[12]} P. Ya. Grozman,
{\it Classification of bilinear invariant operators over tensor fields},
Funct. Anal. Appl., 14 (1980),
58--59.

\item{[13]}  A.A. Kirillov, {\it Infinite dimensional Lie
groups~: their orbits, invariants and representations. The geometry of
moments}, Lect. Notes in Math., 970, 
Springer-Verlag (1982) 101-123.

\item{[14]}  A.A. Kirillov, {\it Orbits of the group of diffeomorphisms  of a circle 
and local superalgebras}, Funct. Anal. Appl., 15:2 (1980)
135-137.

\item{[15]} A.A. Kirillov, {\it K\"ahler structure on $K$-orbits of the group
of diffeomorphisms of a circle}, Funct. Anal. 
Appl. 21:2 (1987) 122-125.

\item{[16]} A.A. Kirillov \& D.V. Yuriev,
{\it K\"ahler geometry of the infinite-dimensional homogeneous space
$M=\Diff_+(S^1)/\hbox{Rot}(S^1)$}, Funct. Anal. 
Appl. 21:4 (1987) 248-294.

\item{[17]} B. Kostant \& S.Sternberg,
{\it The Schwarzian derivative and the conformal geometry of the Lorentz
hyperboloid}, in: Quantum Theories and Geometry (M. Cahen and M. Flato eds.)
Kluwer, 1988, 113-125.

\item{[18]} N.H. Kuiper, {\it Locally projective spaces of dimension
one}, Michigan Math. J., 2 (1954) 95--97.

\item{[19]} V.F. Lazutkin \& T.F. Pankratova, {\it Normal forms
and versal deformations for Hill's equations}, Funct. Anal. 
Appl 9:4 (1975), 306--311.

\item{[20]} P.B.A. Lecomte, P. Mathonet \& E. Tousset,
{\it Comparison of some modules of the Lie algebra of vector fields},
Indag. Math. (N.S.) 7 (1996), no.4, 461-471.

\item{[21]} D.A. Leites, B.A. Feigin,
{\it New Lie superalgebras of string theories},
Group-Theoretic Methods in Physics,
v.1, Moscow (1983) 269-273.

\item{[22]} P. Marcel, V. Ovsienko \& C.Roger,
{\it Extension of the Virasoro and Neveu-Schwarz algebras and
generalized Sturm-Liouville operators}, Lett. Math. Phys. 40
(1997), no.1, 31-39.

\item{[23]} V. Ovsienko,
{\it Classification of third-order linear
differential equations and symplectic sheets of the Gel'fand-Dikii
bracket}, Math. Notes, 47:5 (1990) 465--470.

\item{[24]} O. Ovsienko \& V. Ovsienko, 
{\it Lie derivative of order $n$
on a line. Tensor meaning of the Gelfand-Dickey bracket}, 
Adv. in Soviet Math., 2, 1991. 

\item{[25]} V. Ovsienko, C. Roger, {\it Extension of Virasoro group 
and Virasoro algebra by modules of tensor densities on $S^1$}, 
Funct. Anal. Appl. 30 (1996), no.4, 290-291.

\item{[26]} C. Roger, {\it Extensions centrales d'alg\`ebres et de groupes de
Lie de dimension infinie, alg\`ebre de Virasoro et g\'en\'eralisations},
Rep. Math. Phys. 35 (1995), no.2-3, 225-266.

\item{[27]} G.B. Segal, {\it Unitary representations of some
infinite dimensional groups}, Comm. Math. Phys., 80:3
(1981) 301--342.

\item{[28]} G.B. Segal, {\it The geometry of the KdV equation}
in~: Trieste Conference on topological methods in quantum field theories
- W. Nahm \& al, eds - World Scientific (1990), 96--106.

\item{[29]} E.J. Wilczynski, Projective differential geometry
of curves and ruled surfaces, Leipzig - Teubner, 1906.

\item{[30]} E. Witten, {\it Coadjoint orbits of the Virasoro group},
Comm. Math. Phys., 114:1, (1988) 1--53.

\item{[31]}  P. Marcel, {\it Extensions of the Neveu-Schwarz Lie
superalgebra}, Comm. Math. Phys. 207 (1999), no. 2, 291-306.

\item{[32]}  P. Marcel, {\it Generalized Virasoro algebra and
matrix Sturm-Liouville operators}, J. Geom. Phys., 36  (2000),  no. 3-4, 211--222.

\bye